\documentclass[12pt]{article}

\usepackage{amsmath}
\usepackage{amsthm}
\usepackage{amsfonts}          
\usepackage{amssymb}           
\usepackage[all]{xy}           
\usepackage{cite}

\newlength{\xtrawidth}
\newlength{\xtraheight}
\renewcommand{\baselinestretch}{1.3}
\def\Z{\mathbb{Z}}
\def\fnote#1#2{\begingroup\def\thefootnote{#1}\footnote{#2}
     \addtocounter{footnote}{-1}\endgroup}
\newcommand{\Rep}[1]{\ensuremath{\mathbf{#1}}}
\newcommand{\barRep}[1]{\ensuremath{\overline{\Rep{#1}}}}
\newcommand{\Vt}{{\ensuremath{\widetilde{V}}}}
\newcommand{\Xt}{{\ensuremath{\widetilde{X}}}}
\newcommand{\eref}[1]{eq.~\eqref{#1}}

\newcommand{\cref}[1]{Chapter~\ref{#1}}
\newcommand{\bcenter}{\begin{center}}
\newcommand{\ecenter}{\end{center}}
\newcommand{\beq}{\begin{equation}}
\newcommand{\eeq}{\end{equation}}
\newcommand{\bea}{\begin{eqnarray}}
\newcommand{\eea}{\end{eqnarray}}
\newcommand{\bean}{\begin{eqnarray*}}
\newcommand{\eean}{\end{eqnarray*}}
\newcommand{\ba}{\begin{array}}
\newcommand{\ea}{\end{array}}
\newcommand{\ben}{\begin{enumerate}}
\newcommand{\een}{\end{enumerate}}
\newcommand{\bi}{\begin{itemize}}
\newcommand{\ei}{\end{itemize}}
\newcommand{\bd}{\begin{description}}
\newcommand{\ed}{\end{description}}
\newcommand{\bdiag}{\begin{diagram}}
\newcommand{\ediag}{\end{diagram}}
\def\fnote#1#2{\begingroup\def\thefootnote{#1}\footnote{#2}
     \addtocounter{footnote}{-1}\endgroup}

\def\IC{\mathbb{C}}

\def\IZ{\mathbb{Z}}

\def\IP{\mathbb{P}}

\def\cS{{\mathcal S}}
\def\cO{{\mathcal O}}

\def\cF{{\mathcal F}}

\def\cL{{\mathcal L}}

\def\nn{\nonumber}
\def\td{\mbox{td}}
\def\ch{\mbox{ch}}

\def\tx{\Xt}
\def\av{\wedge^2 V}
\def\atv{\wedge^2 \Vt}
\def\tv{\tilde{V}}

\def\op1{{\mathcal O}_{\IP^1}}

\def\v12{V_1 \otimes V_2}
\def\b{\beta}

\newcommand{\sseq}[3]{0 \longrightarrow #1 \longrightarrow #2 \longrightarrow #3 \longrightarrow 0}
\def\z3z3{\IZ_3 \times \IZ_3}
\def\mssm{SU(3)_C \times SU(2)_L \times U(1)_Y \times U(1)_{B-L}}

\def\dirac{\slash{\! \! \! \! D}}

\begin{document}
\begin{titlepage}  
  \title{
    \hfill{\normalsize  UPR-1104-T} \\[1em]
    {\LARGE A Standard Model from the $E_8 \times E_8$ Heterotic 
      Superstring
      \author{Volker Braun$^{1,2}$, Yang-Hui He$^{1}$, Burt A.~Ovrut$^{1}$,
        and Tony Pantev$^{2}$
        \fnote{~}{vbraun,
          yanghe, ovrut@physics.upenn.edu;
          tpantev@math.upenn.edu}\\[0.5cm]
        {\normalsize $^1$
          Department of Physics,}
        {\normalsize $^2$
          Department of Mathematics}\\
        {\normalsize University of Pennsylvania} \\
        {\normalsize Philadelphia, PA 19104--6395, USA}}
    }
    \date{}
  }
  
  \maketitle

  \begin{abstract}
    In a previous paper, we introduced a heterotic standard model and
    discussed its basic properties.  This vacuum has the spectrum of
    the MSSM with one additional pair of Higgs-Higgs conjugate fields
    and a small number of uncharged moduli. In this paper, the
    requisite vector bundles are formulated; specifically, stable,
    holomorphic bundles with structure group $SU(N)$ on smooth
    Calabi-Yau threefolds with $\z3z3$ fundamental group.  A method
    for computing bundle cohomology is presented and used to evaluate
    the cohomology groups of the standard model bundles. It is shown
    how to determine the $\z3z3$ action on these groups. Finally,
    using an explicit method of ``doublet-triplet splitting'', the
    low-energy particle spectrum is computed.
  \end{abstract}
  \thispagestyle{empty}
\end{titlepage}

\section{Introduction:}

In~\cite{VBphysicsletter}, we presented a standard model within the context 
of the $E_{8} \times E_{8}$ heterotic superstring. These vacua are
$N=1$ supersymmetric and have the following
properties. 
\begin{itemize}
\item The observable sector has gauge group $SU(3)_{C} \times
  SU(2)_{L} \times U(1)_{Y} \times U(1)_{B-L}$, {\it three} families
  of quarks and leptons, each with a {\it right-handed neutrino}, and
  {\it two} Higgs-Higgs conjugate pairs. There is {\it no exotic
    matter}. In addition, there are 6 geometric moduli and a {\it
    small number} of vector bundle moduli. That is, the observable
  sector has exactly the spectrum of the MSSM with one additional
  Higgs-Higgs conjugate pair.
\end{itemize}
Within our context, the visible sector vector bundle is unique. 
All other 
bundles lead to an observable sector spectrum that is not 
realistic, having large numbers of exotic multiplets and Higgs-Higgs 
conjugate pairs. 
\begin{itemize}
\item The structure of the hidden sector depends on whether one
  considers the weakly or strongly coupled regime.  In the strongly
  coupled context, we find a minimal hidden sector with gauge group
  $E_7 \times U(6)$, and {\it no matter fields}. For weak coupling,
  one finds a minimal hidden sector with gauge group $Spin (12)$ and
  {\it two} matter multiplets, each in the $\Rep{12}$ of $Spin(12)$.
  In both cases, there is a {\it small number} of vector bundle
  moduli.
\end{itemize}
There is flexibility in choosing the hidden sector vector bundles
since one can always perform small instanton transitions,
see~\cite{Witten:1995gx, Curio:1998vu, Ovrut:2000qi}. However, those
leading to the minimal spectra just presented are, essentially,
unique.

In~\cite{VBphysicsletter} we presented the basic structure of the
heterotic standard model, but only briefly outlined the requisite
technical results.  The properties of the smooth compactification
manifold, Calabi-Yau threefolds with $\z3z3$ fundamental group, as
well as the action of $\z3z3$ on the associated Wilson lines were
discussed in detail in~\cite{dP9Z3Z3}. However, the construction of
the standard model vacua requires three other ingredients; first,
stable, holomorphic vector bundles with $SU(N)$ structure groups over
this threefold, second, the cohomologies associated with these bundles
and third, the explicit representations of $\z3z3$ on these cohomology
groups.  The low energy spectrum is then identified with the subspace
invariant under the product of these representations with the action
on the Wilson lines. In this paper, we will discuss these three
ingredients in more detail. This will establish the technical basis
for our results in~\cite{VBphysicsletter} and provide the context for
assessing their uniqueness.

The standard model vector bundles are {\it not} constructed from
spectral covers~\cite{FMW, Friedman:1997ih, Andreas:1999ty,
  DonagiPrincipal, Donagi:1998xe, Donagi:1999gc, Curio:2004pf,
  Andreas:2003zb, Diaconescu:1998kg}.  Rather, they are produced using
a generalization of the method of ``bundle extensions'' introduced
in~\cite{Donagi:2000fw, Donagi:2000zs, Donagi:2000zf, Donagi:1999ez,
  Donagi:2003tb, Ovrut:2003zj, Ovrut:2002jk}.  The techniques for
explicitly computing bundle cohomologies, and for finding the
representations of a finite group on the cohomology groups, were
presented~\cite{Donagi:2004su, Donagi:2004ub}. Standard model vacua
require a significant extension of the methods discussed
in~\cite{Donagi:2004qk, Donagi:2004ia}.  Finally, we emphasize that
our computation of the spectrum as the invariant subspace under the
action of $\z3z3$ on the cohomology groups and Wilson lines represents
an explicit method of ``doublet-triplet splitting''
\cite{Green:1987mn, Witten:2001bf}. It is this technique which allows
us to project out all exotic matter and to arrive at the minimal MSSM
spectrum with one additional pair of Higgs-Higgs conjugate fields.

In this paper, we present our computations and discuss the extensive
search that led to the heterotic standard model. However, the full
technical details will be left to future publications~\cite{VBpaper,
  multipletsplitting}. For example, the computation of vector bundle
moduli is more involved than for other fields and will be presented
elsewhere. In this paper, we simply point out that the $\z3z3$
projection greatly reduces the number of such moduli.

\section{Requisite Data:}

We begin with the $E_8 \times E_8$ heterotic string compactified on a
smooth Calabi-Yau threefold $X$. This manifold admits stable, holomorphic 
vector bundles $V$ in the observable $E_8$ sector and $V'$ in the $E_8'$
hidden sector.\footnote{We will distinguish the $E_{8}$ gauge group of the 
hidden sector by denoting it with a prime.} It follows that the 
four-dimensional effective theory will exhibit $N=1$ supersymmetry.

\subsection{The Observable Sector Spectrum}

Consider the minimal supersymmetric standard model, the MSSM.  It is
well-established that neutrinos have a non-vanishing
mass~\cite{Fukuda:1998mi}. Since the MSSM has no exotic multiplets,
$N=1$ supersymmetry will suppress any purely left-handed Majorana
neutrino mass to be too small by several orders of
magnitude~\cite{Langacker:2004xy, Giedt:2005vx}. It
follows that the MSSM must be extended by adding a right-handed
neutrino to each family of quarks/leptons.

We would like to find a vacuum of the $E_8 \times E_8$ heterotic
string whose observable sector is as close to this extended MSSM as
possible. To do this, it is useful to recall that each generation of
quarks/leptons with a right-handed neutrino fits exactly into the
$\Rep{16}$ spin representation of $Spin(10)$. It is compelling,
therefore, to try to spontaneously break the $E_{8}$ gauge group of
the observable sector to $Spin(10)$ as already suggested
in~\cite{Green:1987mn}. This can be accomplished if we choose $V$ to
have structure group $SU(4)$. Then
\begin{equation}\label{b1}
E_8 \longrightarrow Spin(10),
\end{equation}
as desired. With respect to the maximal subgroup $SU(4) \times
Spin(10)$, the adjoint $\Rep{248}$ of $E_8$ decompose as
\begin{equation}
\ba{rcl}
\Rep{248} & \to & \big( \Rep{1},\Rep{45} \big) \oplus \big(
\Rep{15},\Rep{1} \big) \oplus \big( \Rep{4},\Rep{16} \big) \oplus
\big( \barRep{4}, \barRep{16} \big) \oplus \big( \Rep{6},\Rep{10}
\big).
\ea
\end{equation}
The $\big( \Rep{1},\Rep{45} \big)$ contain the gauginos of $Spin(10)$, the
$\big( \Rep{15},\Rep{1} \big)$ correspond to vector bundle moduli
and the remaining representations are the matter fields.

If $X$ is not simply connected, one can introduce, additionally, a 
Wilson line $W$ to further reduce the gauge group. It was shown 
in~\cite{dP9Z3Z3} that to break $Spin(10)$ to a group containing the standard
model gauge group $SU(3)_{C} \times SU(2)_{L} \times U(1)_{Y}$, 
the simplest possibility is to require that $X$ have first fundamental group 
\begin{equation}
\pi_1(X) = \z3z3.
\end{equation}
Calabi-Yau threefolds with this property were explicitly constructed 
in~\cite{dP9Z3Z3}. 
On such manifolds, one can choose  Wilson lines with the 
property that their holonomy group is $hol(W)= \z3z3$.
It was shown in~\cite{dP9Z3Z3} that $W$ will then spontaneously break
\begin{equation}
Spin(10) \longrightarrow \mssm,
\label{b3}
\end{equation}
where, in addition to the standard model gauge group, there is a
gauged $U(1)_{B-L}$ symmetry. With respect to this low energy gauge group, the 
$Spin(10)$ matter fields decompose as
\begin{equation}
  \label{decomp}
  \begin{split}
   \\
    \Rep{16} ~\to &~
    \big(\Rep{3},   \Rep{2}, 1, 1 \big) \oplus
    \big(\barRep{3},\Rep{1}, -4, -1 \big) \oplus
    \big(\barRep{3},\Rep{1}, 2, -1 \big) \oplus
    \big(\Rep{1}, \Rep{2}, -3, -3 \big) \,\oplus
    \\ & \qquad \oplus
     \big(\Rep{1},\Rep{1}, 6, 3 \big) \oplus
    \big(\Rep{1},\Rep{1}, 0, 3 \big),   
    \\
    \Rep{10} ~\to &~
    \big(\Rep{3},\Rep{1}, -2, -2 \big) \oplus
    \big(\barRep{3},\Rep{1}, 2, 2 \big) \oplus
    \big(\Rep{1},\Rep{2}, 3, 0 \big) \oplus
    \big(\Rep{1},\barRep{2}, -3, 0 \big)    
  \end{split}
\end{equation}
where we have displayed the quantum numbers $3Y$ and $3(B-L)$ for
convenience.  The $\barRep{16}$ decomposition is obtained by
conjugation.  We see from \eref{decomp} that the standard model
fermions, including a right-handed neutrino, arise from the
decomposition of the $\Rep{16}$, as expected. Similarly, Higgs
doublets occur in the $\Rep{10}$ of $Spin(10)$.

Note, however, that there may be extra, exotic matter multiplets in the 
spectrum. These include all fields arising from the decomposition of a 
$\barRep{16}$. Additionally, any of the color triplets in the decomposition of
a $\Rep{10}$ are unobserved. Therefore, if one is to be successful in finding 
a heterotic standard model, these exotic matter multiplets must be 
projected out.

\subsection{The Calabi-Yau Threefold $X$}
\label{s:X}

The above discussion implies that one must construct Calabi-Yau
threefolds $X$ with fundamental group $\z3z3$. This was carried out in
detail in~\cite{dP9Z3Z3}. Here, we simply outline those properties of
the construction that are required for the analysis in this paper.

The requisite Calabi-Yau threefolds, $X$, are constructed as
follows. We begin by considering a simply connected Calabi-Yau
threefold, $\Xt$, which is an elliptic fibration over a rational
elliptic surface, $d\mathbb{P}_9$. It was shown in~\cite{dP9Z3Z3} that
there are special $d\mathbb{P}_9$ surfaces which admit a $\z3z3$
action. Furthermore, in a six-dimensional region of moduli space,
$\Xt$ admits an induced $\Z_3 \times \Z_3$ group action which is fixed
point free. The quotient $X= \Xt / (\z3z3)$ is a smooth Calabi-Yau
threefold that is torus-fibered over a singular $d\mathbb{P}_9$ and
has non-trivial fundamental group $\Z_3 \times \Z_3$, as desired.

Specifically, $\Xt$ is a fiber product
\begin{equation} \label{tx} \tx = B_1 \times_{\IP^1} B_2 \end{equation}
of two $d\mathbb{P}_9$ special surfaces $B_1$ and $B_2$. Thus, $\Xt$
is elliptically fibered over both surfaces with the projections
\begin{equation} \pi_1 : \tx \to B_1, \quad \pi_2 : \tx \to B_2. 
\label{b6}
\end{equation}
The surfaces $B_1$ and $B_2$ are themselves elliptically fibered
over $\IP^1$ with maps
\begin{equation}\label{betaB} \b_1 : B_1 \to \IP^1, \quad \b_2 : B_2 \to \IP^1.
\end{equation}

The invariant homology ring of each special $d\mathbb{P}_9$ surface is
generated by two $\z3z3$ invariant curve classes $f$ and $t$ with 
intersections
\begin{equation}
f^{2}=0, \quad ft=3t^{2}.
\label{c1}
\end{equation}
Using projections~\eqref{b6}, these can be lifted to divisor classes  
\begin{equation} \tau_1 = \pi_1^{-1}(t_{1}), \quad \tau_2 = 
\pi_2^{-1}(t_{2}), \quad
\phi = \pi_1^{-1}(f_{1}) = \pi_2^{-1}(f_{2}) \end{equation}
on $\Xt$ satisfying the intersection relations
\begin{equation} 
\phi^2= \tau_1^3= \tau_2^3=0, \quad \phi \tau_1 = 3 \tau_1^2, \quad \phi
\tau_2 = 3 \tau_2^2. \end{equation}
These three classes generate the invariant homology ring of $\Xt$. For example,
\begin{equation}
span_{\mathbb{C}} \{\phi,\tau_{1},\tau_{2}\}= H_{4}(\Xt, \mathbb{C})^{\z3z3} 
\simeq H^{1,1}(X).
\label{cc1}
\end{equation}
It follows that $h^{1,1}(X)=3$. Similarly, one can show that 
$h^{1,2}(X)=3$. Hence, $X$ has six geometric moduli; three 
Kahler moduli and three complex structure moduli.
Finally, the Chern classes of $\Xt$ can be shown to be
\begin{equation}\label{chernX}
    c_1(T\Xt) = c_3(T\Xt) = 0, \quad
    c_2(T\Xt) = 12(\tau_1^2 + \tau_2^2).
\end{equation}

\subsection{The Holomorphic $SU(4)$ Bundle $V$}

Next, we produce the requisite observable sector 
bundles $V$ on $X$. This is accomplished by
constructing stable, holomorphic vector bundles $\Vt$
with structure group $SU(4)$ over $\Xt$ that are equivariant
under the action of $\Z_3 \times \Z_3$. Then $V=\Vt/(\z3z3)$.

The vector bundles $\Vt$ are constructed using a generalization of the
method of ``bundle extensions'' introduced in~\cite{Donagi:2000fw,
  Donagi:2000zs, Donagi:2000zf, Donagi:1999ez, Donagi:2003tb,
  Ovrut:2003zj, Ovrut:2002jk}. Specifically, $\Vt$ is the extension
\begin{equation}\label{V12} \sseq{V_1}{\Vt}{V_2} \end{equation}
of two rank two bundles $V_1$ and $V_2$ on $\Xt$. These bundles
are of the form
\begin{equation} V_i = \cL_i \otimes \pi_2^*W_i, \quad i=1,2 
\label{d3}
\end{equation}
for some line bundles $\cL_i$ on $\Xt$ and rank 2 bundles $W_i$ on
$B_2$. The rank two bundles $W_{i}$ are themselves extensions
\begin{equation} \sseq{\cO_{B_2}(a_{i} f_{2})}{W_i}{\cO_{B_2}(b_{i} f_{2}) 
\otimes I_{k_{i}}}, 
\label{d1} 
\end{equation}
where $a_{i},b_{i}$ are integers and $I_{k_{i}}$ is the ideal sheaf of some
$k_{i}$-tuple of points on $B_2$. That is,~\eqref{d1} gives us a
prescription to build rank two bundles on $B_2$,~\eqref{d3} to produce
two rank two bundles on $\Xt$ and, finally, we use~\eqref{V12} to construct 
$\Vt$.

One must specify not only the bundles $\Vt$, but their transformations
under $\z3z3$ as well. To do this, first notice that for the $\z3z3$
action on the space of extensions to be well-defined, the line bundles
$\cO_{B_{2}}(a_{i}f_{2})$, $\cO_{B_{2}}(b_{i}f_{2})$ and $\cL_{i}$
must be equivariant under the finite group action. In this case, the
space of extensions will carry a representation of $\z3z3$. An
equivariant rank four vector bundle will be any $\Vt$ that does not
transform under this action.  A $\Vt$ with this property will inherit
an explicit equivariant structure from the action of $\z3z3$ on its
constituent line bundles.  Having found such a $\Vt$, one can
construct $V=\Vt/(\z3z3)$ on $X$, as required.

To proceed, therefore, one must consider the action of $\z3z3$ on line bundles
and show how to construct line bundles that are equivariant. Two natural 
one-dimensional representations of $\z3z3$ are defined by
\begin{equation}
{\chi}_{1}(g_{1})=\omega,\quad
{\chi}_{1}(g_{2})=1; \qquad
{\chi}_{2}(g_{1})=1, \quad
{\chi}_{2}(g_{2})=\omega,
\label{e1}
\end{equation}
where $g_{1,2}$ are the generators of the two $\mathbb{Z}_{3}$ factors, 
${\chi}_{1,2}$ are two group characters of $\z3z3$ and $\omega = 
e^{\frac{2 \pi i}{3}}$ is a third root of unity. All other one-dimensional 
representations are products of~\eqref{e1} 
and, in any case, do not appear in our construction. 
Note that none of these representations is faithful.

%
%
Let us consider an explicit example of a $\z3z3$ action on a line bundle.
Recall that $\cO_{\Xt} \simeq \Xt \times \IC \ni (p,v)$ is the
trivial line bundle on $\Xt$. The simplest action of a group element $g \in
\z3z3$ on $\cO_{\Xt}$ is by translation of $p$ to $g(p)$, with no
action on $v$. However, for any representation $\chi$ we can define a twisted 
action on $\cO_{\Xt}$ by
\begin{equation}
    (p,v) \mapsto (g(p), \chi(g) v).
\end{equation}
In this paper, we denote $\cO_{\Xt}$ carrying this twisted action by
$\chi \cO_{\Xt}$. It is straightforward to show that $\chi \cO_{\Xt}$
is equivariant under $\z3z3$, as desired. 
We may similarly define line bundles $\chi \cL$ on $\Xt$, $\chi
\cO_{B_i}(nf)$ on $B_i$ and $\chi \op1(n)$ on $\IP^1$.

Finally, having constructed equivariant holomorphic bundles 
$\Vt$ with structure group $SU(4)$ over $\Xt$, one must ensure that they are 
stable. For an arbitrary holomorphic vector bundle $\cF$, a complete 
proof of stability is extremely complicated. However, one can show that 
a bundle $\cF$ with vanishing first Chern class is stable only if
\begin{equation}\label{stability}
    H^0(\Xt, \cF) = H^0(\Xt, \cF^*) = 0, \quad H^0(\Xt, \cF \otimes
    \cF^*) = 1.
\end{equation}
\label{e3}
\noindent We will use these criteria as highly non-trivial checks on the 
stability of $\Vt$, as well as on the hidden sector bundle $\Vt'$.

\subsection{Computing the Particle Spectrum}

A method for computing the low-energy particle spectrum after
compactification on $X$ with a holomorphic vector bundle $V$ and
possible Wilson line $W$ was presented in~\cite{Donagi:2004su,
  Donagi:2004ub, Candelas:1985en, Sen:1985eb, Witten:1985xc,
  Breit:1985ud, Greene:1986ar, Greene:1986jb, Aspinwall:1987cn} and
will be used in this paper. The spectrum is identified with the zero
modes of the Dirac operator on $X$ ``twisted'' by the bundle $V \oplus
W$. The zero modes are the invariant elements of certain bundle
cohomology groups.  In this method, one does not actually make use of
$X$ and $V$, all computations being performed on the covering space
$\Xt$ with bundle $\Vt$.

To be specific, let us consider the observable sector discussed above.
In this case, $\Vt$ has structure group $SU(4)$ which breaks $E_{8}$ to 
$Spin(10)$. Furthermore, $\Xt$ admits a free $\z3z3$ action and $\Vt$ 
is equivariant under this action. Let
$R$ be a representation of $Spin(10)$ and denote the associated
$\Vt$ bundle by $U_{R}(\Vt)$. One first constructs 
$H^1( \Xt,U_{R}(\Vt))$ for all non-trivial bundles $U_{R}(\Vt)$.
When $U_{R}(\Vt)$ is trivial, one considers $H^0( \Xt, \cO_{\tx})$ which is
always one-dimensional and carries the trivial representation of $\z3z3$.
The next step is to find the representation of $\Z_3
\times \Z_3$ on
$H^1( \Xt,U_{R}(\Vt))$. Choosing $\Vt$ to be equivariant guarantees
that these actions exit. Finally,  tensor each such representation with
the action of the Wilson line on $R$. The zero mode
spectrum is then the invariant subspace under this joint group
action. In summary, the particle spectrum is
\begin{multline}
  \label{spec} 
  \ker(\dirac_{\Vt}) = 
\left( H^0(\tx, \cO_{\tx}) \otimes \Rep{45} \right)^{\z3z3}  
\oplus 
  \left( H^1\big(\tx, {\rm ad}(\tv) \big) 
         \otimes \Rep{1} \right)^{\z3z3} 
\oplus \\ \oplus 
  \left( H^1(\tx, \tv) \otimes \Rep{16} \right)^{\z3z3}  
\oplus
  \left( H^1(\tx, \tv^*) \otimes \barRep{16} \right)^{\z3z3} 
\oplus 
  \left( H^1(\tx, \atv ) \otimes \Rep{10} \right)^{\z3z3}
  \,,
\end{multline}
where the superscript indicates the $\z3z3$ invariant subspace.

Although we have illustrated our method for the observable sector, it is 
completely general, applying to the hidden sector as well. It 
follows that the computation of cohomology groups, and the $\z3z3$ action 
on these groups, is a major ingredient of our construction.

\subsection{Physical Constraints}

Obtaining realistic particle physics in the observable sector 
requires the the following additional constraints on $\Vt$.
\begin{description}
\item[1. Three Generations:] To ensure that there are three
  generations of quarks and leptons in the low-energy spectrum, one
  must require that
  \begin{equation}\label{3fam1}
    h^1(X, V) - h^1(X, V^*) = 3.
  \end{equation}
  Using Serre duality, and assuming $\Vt$ satisfies \eref{stability},
  the Atiyah-Singer index theorem implies
  \begin{equation}
    -h^1(X, V) + h^1(X, V^*) = \int_X \ch(V) \td(TX) = \frac12 \int_X
    c_3(V) = -3.
    \label{burt22}
  \end{equation}
  Therefore, one must demand $c_{3}(V)=-6$ or, equivalently, that
  \begin{equation}\label{3fam}
    c_{3}\big( \Vt \big)= -6 \times |\z3z3| = -54 .
  \end{equation}
\item[2. No Exotic Matter in the Observable Sector:] The previous
  constraint ensures that there are precisely three chiral generations
  descending from the $\Rep{16}$ representations.  However, there
  remains, in general, a large number of additional low energy
  multiplets which descend from vector-like $\barRep{16}-\Rep{16}$
  pairs. These ``exotic multiplets'' are unobserved. Therefore, we
  place a very strong restriction on $\Vt$ and demand that there be no
  exotic multiplets in its low-energy spectrum. Referring
  to~\eqref{spec}, we see that the simplest way to ensure this is to
  require
  \begin{equation}\label{noexotic}
    h^1( \Xt, \Vt^*)=0.
  \end{equation}
  To our knowledge, this has never been accomplished in any other
  phenomenological string vacua. These typically have exotic
  multiplets in vector-like pairs which, it is hoped, acquire heavy
  masses. In our work, we constrain our spectrum to be as close to the
  MSSM as possible.
\item[3. Small Number of Higgs Doublets:] The number of $\Rep{10}$
  zero modes is given by $h^1(\Xt,{\wedge}^{2}\Vt)$. Since the Higgs
  fields arise from the decomposition of the $\Rep{10}$, we must not
  set the associated cohomology to zero. Rather, we restrict $\Vt$ so
  that
  \begin{equation}
    h^1(\Xt,{\wedge}^{2}\Vt) \quad  \mbox{is minimal},
    \label{f1}
  \end{equation}
  but non-vanishing.  
\item[4. Doublet-Triplet Splitting:] Inspecting \eqref{decomp}, we see
  that the decomposition of the $\Rep{10}$ representation contains, in
  addition to Higgs fields, unwanted ``exotic'' color triplet
  multiplets. We require, therefore, that
  $(H^1(\Xt,{\wedge}^{2}\Vt)\otimes \Rep{10})^{\z3z3}$ contain only
  the Higgs-doublets $\big(\Rep{1},\Rep{2}, 3, 0 \big) \oplus
  \big(\Rep{1},\Rep{2}, -3, 0 \big)$, thus projecting out the color
  triplets at low energy.  This provides a natural solution to the
  doublet-triplet splitting problem.  Note that this mechanism is not
  confined to doublets/triplets. It applies to the components of any
  multiplet, greatly reducing the spectrum after taking the $\z3z3$
  quotient.
\end{description}

The vector bundle $\Vt'$ of the hidden sector must also obey the 
following constraint.

\begin{description}
\item[5. Anomaly Cancellation:] For the theory to be consistent, one
  must require the cancellation of all anomalies. Through the
  Green-Schwarz mechanism, this requirement relates the observable and
  hidden sector bundles, imposing the constraint on the second Chern
  classes that
  \begin{equation}\label{eff}
    [\mathcal{W}_5]= c_2\big( T\Xt \big) - c_2\big( \Vt  \big) -
    c_2\big( \Vt' \big)
  \end{equation}
  be an effective class. In the strongly coupled heterotic string,
  $[\mathcal{W}_5]$ is the class of the holomorphic curve around which
  a bulk space five-brane is wrapped. In the weakly coupled case,
  $[\mathcal{W}_5]$ must vanish. In either case, $c_{2}(T\Xt)$ and
  $c_{2}(\Vt)$ are fixed by previous considerations.
  Therefore,~\eqref{eff} becomes a constraint on the second Chern
  class of $\Vt'$.
\end{description}

\section{The Solution:}\label{s:sol}

In this section, explicit bundles $\Vt$ and $\Vt'$ satisfying the 
requisite data outlined above are constructed. We begin by considering the 
observable sector bundle.

\subsection{The Observable Sector Bundle $\Vt$}

After an extensive search, we found a unique solution for $\Vt$ that is
compatible with all of our constraints. It is constructed as follows. First
consider the rank two bundles $W_{i}$ for $i=1,2$ on $B_{2}$.
Take $W_{1}$ to be
\begin{equation}
W_{1}=\cO_{B_2} \oplus \cO_{B_2}.
\label{f11}
\end{equation}
Note that this is the trivial extension of~\eqref{d1} with 
$a_{1}=b_{1}=k_{1}=0$. Now let
$W_{2}$ be an equivariant bundle in the extension space of
\begin{equation}
    \sseq{\cO_{B_2}(-2f_{2})}{W_{2}}{\chi_2\cO_{B_2}(2f_{2}) \otimes I_9},
\label{DD}
\end{equation}
where for the ideal sheaf $I_9$ of 9 points we take a generic
$\z3z3$ orbit. Second, choose the two line bundles $\cL_{i}$ for
$i=1,2$ on $\Xt$ to be
\begin{equation}
\cL_{1}= \chi_2 \cO_{\Xt}(-\tau_1 + \tau_2)
\label{f2}
\end{equation}
and
\begin{equation}
\cL_{2}= \cO_{\Xt}(\tau_1 - \tau_2)
\label{f3}
\end{equation}
respectively. Then, the two rank 2 bundles $V_{1,2}$ defined 
in~\eref{d3} are given by
\begin{eqnarray}\label{solV}
V_1 &=& \chi_2 \cO_{\Xt}(-\tau_1 + \tau_2) \oplus \chi_2 \cO_{\Xt}(-\tau_1 + \tau_2) \nn \\
V_2 &=& \cO_{\Xt}(\tau_1 - \tau_2) \otimes \pi_2^*W_{2}.
\end{eqnarray}
Note that $V_1$ is of a special form, having no ideal sheaves and
being itself the trivial extension, namely, a direct sum of two line
bundles. The observable sector bundle $\Vt$ is then defined as an equivariant element of the 
space of extensions~\eqref{V12}. We now show that $\Vt$, so-defined, satisfies all of the 
requisite constraints.

Let us begin with the three generation condition. Computing the Chern classes of $\Vt$,
we find that
\begin{equation}\label{chernVt}
    c_1(\Vt)=0, \quad
    c_2(\Vt)=-2\tau_1^2 + 7 \tau_2^2 + 4\tau_1\tau_2, \quad
    c_3(\Vt)=-54.
\end{equation}
Note that $c_3(\Vt)=-54$, as required by the three generation condition~\eqref{3fam}.

To count the number of exotic multiplets in the observable sector, it follows from~\eqref{spec}
that one must compute $h^{1}(\Xt, \Vt^{*})$. We find it more convenient to calculate
$h^{2}(\Xt, \Vt)$ and then use Serre duality to find $h^{1}(\Xt, \Vt^{*})$. Furthermore, to discuss
the stability of $\Vt$ as well as the number of $\Rep{16}$ representations, we see 
from~\eqref{stability} and~\eqref{spec} that we need to know $h^{i}(\Xt, \Vt)$ for $i=0,1,3$ as well.
To do this, recall that $\Vt$ is in the short exact bundle sequence~\eqref{V12}. This induces 
a long exact sequence involving the desired cohomology groups  $H^{i}(\Xt, \Vt)$ 
for $i=0,1,2,3$. These groups can be calculated if we can compute the adjacent terms in the
long exact sequence, namely $H^{i}(\Xt, \cF)$ where $\cF=V_{1}$ and $V_{2}$. 
This can indeed be accomplished
using Leray spectral sequences.  Exploiting the fact that $\Xt$ is ``doubly'' elliptic, with $\pi_{i}$ 
in~\eqref{b6} projecting $\Xt$ to $B_{i}$ and $\beta_{i}$ in~\eqref{betaB}
mapping $B_{i}$ to $\IP^1$, the spectral sequence for any sheaf $\cF$ simplifies to
\begin{equation}\label{leray0}
H^0(\Xt, \cF) = H^0(\IP^1, \b_{i*} \pi_{i*} \cF)
\end{equation}
and 
\begin{equation}
  \label{leray1}
  \vcenter{\xymatrix@=6mm{
    & 0 \ar[d] \\
    & *+<0mm>{\strut}\save[]*{ 
      H^0(\IP^1, R^1\b_{i*}\pi_{i*}\cF) } \restore \ar[d] &  
    & *+<0mm>{\strut}\save[]*{ 
      H^0(\IP^1, \b_{i*} R^1\pi_{i*}\cF) } \restore \ar@{=}[d] \\
    0 \ar[r] & H^1(B_i, \pi_{i*}\cF) \ar[d] \ar[r] 
    & *+[F]{H^1(\Xt, \cF)} \ar[r] 
    & H^0(B_i, R^1\pi_{i*}\cF) \ar[r] 
    & H^2(B_i,\pi_{i*}\cF) \ar[r] 
    & \cdots \\
    & *+<0mm>{\strut}\save[]*{ 
      H^1(\IP^1, \b_{i*}\pi_{i*}\cF) } \restore \ar[d] \\
    & 0
  }}
\end{equation}
where we have boxed the term we wish to compute in \eqref{leray1}. By
$R^1\pi_*$ and $R^1\b_*$ we mean the first higher images of the
push-down maps $\pi_*$ and $\b_*$ respectively. To calculate the cohomology 
spaces $H^i$ for $i=2,3$ one can simply use Serre duality which, 
on a Calabi-Yau threefold,
$\Xt$ states that
\begin{equation} \label{serre} H^i(\Xt, \cF) \simeq H^{3-i}(\Xt, \cF^*)^*, \quad
i=0,1,2,3.
\end{equation}

Equations~\eqref{leray0}
and~\eqref{leray1} reduce the computation of $H^{i}(\Xt, \cF)$ for $i=0,1,2,3$
to the evaluation of certain cohomology spaces on $\IP^{1}$. 
In the present case, $\cF=V_{1}, V_{2}$. 
Using~\eqref{solV} and the expressions for the push-downs given by
\begin{equation}
  \label{pushdown}
  \begin{gathered}
  \cO_{B_i}(nf) = \beta_i^\ast \cO_{\IP^1}(n) \,,~ n\in \Z \\
  \begin{aligned}
    \beta_{i\ast} \cO_{B_i}(2t)  &= 6 \cO_{\IP^1} \qquad
    &
    \beta_{i\ast} \cO_{B_i}(-2t) &= 0 \qquad
    &
    R^1\beta_{i\ast} \cO_{B_i}(2t)  &= 0
    \\
    \beta_{i\ast} \cO_{B_i}(t)   &= 3 \cO_{\IP^1}
    &
    \beta_{i\ast} \cO_{B_i}(-t)  &= 0
    &
    R^1\beta_{i\ast} \cO_{B_i}(t)   &= 0
\end{aligned} \\
    \begin{aligned}
    R^1\beta_{1\ast} \cO_{B_1}(-t)  &= 3 \chi_1 \cO_{\IP^1}(-1) \qquad
    &
    R^1\beta_{1\ast} \cO_{B_1}(-2t) &= 6 \chi_1 \cO_{\IP^1}(-1) \qquad
    \\
    R^1\beta_{2\ast} \cO_{B_2}(-t)  &= 3 \cO_{\IP^1}(-1)
    &
    R^1\beta_{2\ast} \cO_{B_2}(-2t) &= 6 \cO_{\IP^1}(-1)
    \,, 
  \end{aligned}
\end{gathered}
\end{equation}
the cohomology spaces on $\IP^1$ can easily be computed.

Putting everything together, we find that
\begin{eqnarray}\label{hVt}
  h^0(\Xt, \Vt) &=& h^3(\Xt, \Vt^*) = 0 \nn \\
  h^1(\Xt, \Vt) &=& h^2(\Xt, \Vt^*) = 27 \nn \\
  h^2(\Xt, \Vt) &=& h^1(\Xt, \Vt^*) = 0 \nn \\
  h^3(\Xt, \Vt) &=& h^0(\Xt, \Vt^*) = 0.
\end{eqnarray}
Note that these results are consistent with equation~\eqref{noexotic} 
for the absence of exotic multiplets arising from vector-like $\barRep{16}
-\Rep{16}$ pairs. They also satisfy the necessary conditions, given 
in~\eqref{stability}, for $\Vt$ to be a stable bundle. Finally,  
cohomology~\eqref{hVt} is consistent with the Atiyah-Singer index theorem 
for $\Vt$ on $\Xt$ and the three generation condition~\eqref{3fam}.

Next, consider the $\Rep{10}$ representations of $Spin(10)$ which,
from~ \eqref{decomp}, give rise to Higgs doublets. It follows from~\eqref{f1} 
that one must compute $h^1(\Xt,\atv)$ and show it to be minimal, 
but non-vanishing. To do this, 
note that $\atv$ lies in the intertwined sequences
%
\begin{equation}
  \vcenter{\xymatrix@R=5mm{
      &&& 0 \ar[d] \\
      &&& \av_2 \ar[d]\\ 
      0 \ar[r] & \av_1 \ar[r] & \atv \ar[r] & Q \ar[d] \ar[r] & 0 \,, \\
      &&& V_1 \otimes V_2 \ar[d] \\
      &&& 0  
    }}
\label{gg2}
\end{equation}
where $Q$ is the quotient of the map $\av_2 \to \atv$. Since
$V_{1,2}$ are rank 2, $\av_{1,2}$ are line bundles and, using~\eqref{solV}, 
are given by
\begin{equation}
    \av_1 = \chi_2^2 \cO_{\Xt}(-2\tau_1+2\tau_2), \quad
    \av_2 = \cO_{\Xt}(2\tau_1-2\tau_2).
\end{equation}
The bundle sequences~\eqref{gg2} give rise to two long exact cohomology 
sequences. To compute $H^i(\Xt,\atv)$ for $i=0,1,2,3$, one must 
compute the adjacent terms in these sequences, namely, $H^i(\Xt,\cF)$
for $\cF= \av_1$, $\av_2$ and $\v12$. This can be accomplished using the 
Leray spectral sequences given in~\eqref{leray0} and ~\eqref{leray1}. We find, 
happily, that the entire cohomology of both $\av_{1,2}$
vanish. It follows that
\begin{equation}
    H^i(\Xt,\atv) \simeq H^i(\Xt,\v12), \quad i=0,1,2,3.
\label{AA}
\end{equation}
Finally, setting $\cF=\v12$ in~\eqref{leray0} and~\eqref{leray1}, we find
\begin{equation}
h^0(\Xt,\atv)=h^3(\Xt,\atv)=0, \quad h^1(\Xt,\atv)=h^2(\Xt,\atv)=14.
\label{gg3}
\end{equation}
Although not immediately apparent, an exhaustive search reveals that
\begin{equation}
h^1(\Xt,\atv)=14
\label{hh1}
\end{equation}
is the minimal number of $\Rep{10}$ representations 
within our context.

As discussed previously, knowledge of the bundle cohomology 
groups corresponding to the $\Rep{16}$ and $\Rep{10}$ representations 
is not sufficient to determine the low energy spectrum. One must also 
evaluate the explicit action of $\z3z3$ on these spaces. 
First consider the cohomology space $H^{1}(\Xt, \Vt)$ associated 
with the $\Rep{16}$ representation. In this case, one can determine 
the $\z3z3$ action using a simple argument. Note from~\eqref{hVt} that 
\begin{equation}
h^{1}(\Xt,\Vt)=27.
\label{burt33}
\end{equation}
Furthermore,~\eqref{hVt} specifies that $h^{1}(\Xt,\Vt^{*})$ vanishes and, 
hence, $h^{1}(X,V^{*})=0$. Then~\eqref{burt22} becomes
\begin{equation}
h^{1}(X,V)=3.
\label{burt11}
\end{equation}
Comparing~\eqref{burt33} to~\eqref{burt11}, it follows that the invariant 
subspace of the $\z3z3$ action on $H^{1}(\Xt,\Vt)$ must be three-dimensional.
That is,
\begin{equation}
h^{1}(\Xt,\Vt)^{\z3z3}=3.
\label{burt44}
\end{equation}
Now, $\Vt$ is equivariant under the explicit action of $\z3z3$ 
discussed earlier. However, as far as cohomology is concerned, 
one can consider nine equivariant actions specified by the characters
${\chi_{1}}^{p}{\chi_{2}}^{q}$ for $p,q=0,1,2$, on $\Vt$. Since the 
bundle is the same, $H^{1}(\Xt,\Vt)$,~\eqref{burt33} 
and~\eqref{burt11} remain unchanged. However, the 
action of $\z3z3$ on $H^{1}(\Xt,\Vt)$ will be altered for each choice of
${\chi_{1}}^{p}{\chi_{2}}^{q}$. Specifically, the original representation 
will be multiplied by the character. Since~\eqref{burt33} 
and~\eqref{burt11} remain unchanged, we conclude that
\begin{equation}
h^{1}(\Xt, {\chi_{1}}^{p}{\chi_{2}}^{q}\Vt)^{\z3z3}=3
\label{burt55}
\end{equation}
for each choice of $p,q=0,1,2$. The only way this can be true is if the 
original $\z3z3$ action is
\begin{equation}
H^{1}(\Xt,\Vt)={\rm Reg}(\z3z3)^{\oplus 3},
\label{gg5}
\end{equation}
where the regular representation of $\z3z3$ is given by
\begin{equation}
{\rm Reg}(\z3z3)=1 \oplus \chi_{1} \oplus \chi_{2} \oplus {\chi_{1}}^{2}
\oplus \chi_{1} \chi_{2} \oplus {\chi_{2}}^{2} \oplus {\chi_{1}}^{2} \chi_{2}
\oplus \chi_{1} {\chi_{2}}^{2} \oplus {\chi_{1}}^{2} {\chi_{2}}^{2}.
\label{gg6}
\end{equation}
Note that~\eqref{gg6} contains all of the irreducible representations 
of $\z3z3$.

Now consider the the cohomology space $H^1(\Xt, \wedge^2\Vt )$ 
associated with the $\Rep{10}$ representation. We know 
from~\eqref{hh1} that $h^1(\Xt, \wedge^2\Vt )=14$. One can find 
the $\z3z3$ action on this space as follows. Recall from~\eqref{AA} that
\begin{equation}
    H^1(\Xt,\atv) \simeq H^1(\Xt,\v12).
\label{BB}
\end{equation}
It follows from~\eqref{solV} that
\begin{equation}
\v12=(\pi_{2}^{*}(\chi_{2}W_{2}))^{\oplus2},
\label{CC}
\end{equation}
where $W_{2}$ is defined by~\eqref{DD}. 
Note that the $\chi_{2}$ action on the line bundles in~\eqref{solV} modifies 
the equivariant structure of $W_{2}$, which we indicate by $\chi_{2} W_{2}$.
Then
\begin{equation}
    H^1(\Xt,\v12) \simeq  H^1(\Xt,\pi_{2}^{*}(\chi_{2}W_{2}))^{\oplus2}.
\label{EE}
\end{equation}
For ease of notation, we will, henceforth, denote $\chi_{2}W_{2}$
simply as $W_{2}$.
To proceed, one must calculate $H^1(\Xt,\pi_{2}^{*}W_{2})$. This can 
be accomplished using~\eqref{leray1} with $i=2$ and 
$\cF=\pi_{2}^{*}W_{2}$, as well as the push-down formulas
\begin{eqnarray}\label{pushV12}
\b_{2*}W_{2} &=& \chi_{2}\op1(-2) \oplus {\chi_2}^{2} \op1(-1), \nn\\
R^1\b_{2*}W_{2} &=& {\chi_2}^{2} \op1(1) \oplus \chi_{2}\op1 \oplus
        \bigoplus_{i=1}^3 \cO_{\b_2(p_k)},
\end{eqnarray}
where $p_k$ are points in $B_2$ associated with the ideal sheaf $I_9$
in the definition of $W_{2}$. Using~\eqref{pushV12}, we find that the terms 
adjacent to  $H^1(\Xt,\pi_{2}^{*}W_{2})$ in~\eqref{leray1} are
\begin{equation}
H^0(\IP^1, \beta_{2*}R^1\pi_{2*}(\pi_{2}^{*}W_{2})) = 0
\label{FF}
\end{equation}
and
\begin{eqnarray}
H^1(\IP^1, \b_{2*}W_{2}) &=& \chi_1^2 \chi_2
 \nn \\
H^0(\IP^1,  R^1\b_{2*}W_{2}) &=&
        (\chi_2^2 \oplus \chi_1 \chi_2^2) \oplus \chi_2 \oplus
                (1 \oplus \chi_1 \oplus \chi_1^2).
\label{GG}
\end{eqnarray}                                                          
Expression~\eqref{FF} cuts off the horizontal sequence in~\eqref{leray1}, 
yielding
\begin{equation}
H^1(\Xt,\pi_{2}^{*}W_{2}) \simeq H^{1}(B_{2}, W_{2}).
\label{HHH}
\end{equation}
On the other hand,~\eqref{GG} inserted into the vertical sequence 
of~\eqref{leray1} implies, using~\eqref{HHH}, that
\begin{equation}
H^1(\Xt, \pi_2^* W_2) =  1 \oplus  \chi_{1} \oplus \chi_2 \oplus  
\chi_1^2 \oplus \chi_2^2 \oplus  \chi_1 \chi_2^2 \oplus 
\chi_1^2 \chi_2.
\label{II}
\end{equation}                                                          
Putting~\eqref{BB},~\eqref{EE} and~\eqref{II} together, we find that the
$\z3z3$ action on $H^1(\Xt, \wedge^2\Vt )$ is
\begin{equation}
H^1(\Xt, \wedge^2\Vt )=2 \oplus 2\chi_1 \oplus 2\chi_2 \oplus
2\chi_1^2 \oplus 2\chi_2^2 \oplus 2\chi_1\chi_2^2 
\oplus 2\chi_1^2\chi_2.
\label{gg7}
\end{equation}
%


Having determined $\Vt$, the cohomology groups $H^{i}(\Xt, U_{R}(\Vt))$
and the action of $\z3z3$ on these spaces, it remains to compute the 
low energy spectrum of the observable sector. To do this, one must give the 
representation of $hol(W)=\z3z3$ on each multiplet $R$. We can choose the 
Wilson line $W$ to have the following actions. 

First consider $R=\Rep{16}$. Then
\begin{equation}
\Rep{16}=\left(\chi_{1}^{2} \chi_{2}(\Rep{3},\Rep{2}) \oplus \chi_{1}^{2}
\chi_{2}^{2}(\barRep{3}, \Rep{1}) \oplus \chi_{1}^{2}
(\Rep{1}, \Rep{1})\right) \oplus \left(\chi_{2}^{2}
(\barRep{3}, \Rep{1}) \oplus (\Rep{1}, \Rep{2})\right)
\oplus (\Rep{1}, \Rep{1}).
\label{r2}
\end{equation}
The terms are grouped according to the $\Rep{10} \oplus \barRep{5} \oplus
\Rep{1}$ decomposition of $\Rep{16}$ under $SU(5)$. For simplicity, 
we have only given the $SU(3)_{C} \times SU(2)_{L}$ quantum numbers
in~\eqref{r2}. Tensoring this with the action~\eqref{gg5},~\eqref{gg6}
of $\z3z3$ on $H^{1}(\Xt, \Vt)$, we find that the invariant 
subspace is spanned by three families of quarks/leptons, each family 
transforming as
\begin{equation}
\big(\Rep{3},   \Rep{2}, 1, 1 \big) \,,\quad
\big(\barRep{3},\Rep{1}, -4, -1 \big) \,,\quad
\big(\barRep{3},\Rep{1}, 2, -1 \big)
\label{15}
\end{equation}
and
\begin{equation}
\big(\Rep{1},\Rep{2}, -3, -3 \big) \,,\quad
\big(\Rep{1},\Rep{1}, 6, 3 \big) \,,\quad
\big(\Rep{1},\Rep{1}, 0, 3 \big)
\label{16}
\end{equation}
under $SU(3)_{C} \times SU(2)_{L} \times U(1)_{Y} \times U(1)_{B-L}$.
We have displayed the quantum numbers $3Y$ and $3(B-L)$ for
convenience. Note from eq.~\eqref{16} that each family contains a
right-handed neutrino, as desired.

Now consider $R=\Rep{10}$. We find that
\begin{equation}
\Rep{10}= 
\left(\chi_{1}^{2}(\Rep{1},\Rep{2}) \oplus \chi_{1}^{2}\chi_{2}^{2}
(\Rep{3}, \Rep{1}) \right) \oplus 
\left(\chi_{1}(\Rep{1}, \barRep{2}) \oplus \chi_{1}\chi_{2}
(\barRep{3}, \Rep{1}) \right).
\label{20}
\end{equation}
where we have grouped the terms in the $\Rep{5} \oplus \barRep{5}$
decomposition of $\Rep{10}$ under $SU(5)$.
Tensoring this with the the action~\eqref{gg7} of $\z3z3$ on 
$H^1(\Xt, \wedge^2\Vt )$, one finds that the
invariant subspace consists of $\it two$ copies of the vector-like
pair
\begin{equation}
\left( \Rep{1},\Rep{2}, 3, 0 \right) \,,\quad
\left( \Rep{1},\barRep{2}, -3,  0 \right)
\,.
\label{21}
\end{equation}
That is, there are two Higgs-Higgs conjugate pairs occurring as zero
modes in the observable sector. Note that the unobserved color triplet 
multiplets have been projected out, as desired. This is an explicit 
mechanism for ``doublet-triplet'' splitting.

We conclude that the zero mode spectrum of the observable
sector 1) has gauge group $SU(3)_{C} \times SU(2)_{L} \times U(1)_{Y}
\times U(1)_{B-L}$, 2) contains {\it three families} of quarks and
leptons each with a {\it right-handed neutrino}, 3) has {\it two}
Higgs-Higgs conjugate pairs and 4) {\it contains no exotic fields} of  
of any kind. Additionally, there are 5) a
{\it small number} of uncharged vector bundle moduli. These arise
from the invariant subspace of $H^1\big(\Xt, \Vt \otimes
\Vt^*\big)$ under the action of $\Z_3 \times \Z_3$ and will be computed 
elsewhere.

\subsection{The Hidden Sector Bundle $\Vt'$}
The vacuum also contains a stable, holomorphic vector bundle,
$V'$, on $X$ whose structure group is in the $E_8'$ of the hidden
sector. Additionally, there can be a Wilson line $W'$ on $X$ whose $\z3z3$
holonomy group is contained in $E_{8}'$. However, to allow for
spontaneously breaking of the $N=1$ supersymmetry via gaugino 
condensation in the 
hidden sector, it is expedient to reduce $E_{8}'$ as little as possible.
With this in mind, we will choose $W'$ to be trivial. 

As for $V$, we construct $V'$ by building stable,
holomorphic vector bundles $\Vt'$ over $\Xt$ which are equivariant
under $\Z_3 \times \Z_3$ using the method of ``bundle extensions''.
$V'$ is then obtained as the quotient of $\Vt'$ by $\Z_3
\times \Z_3$. This bundle must satisfy the anomaly cancellation
condition \eqref{eff}. The simplest possibility is that $\Vt'$ is the
trivial bundle. However, in this case, we find that $[\mathcal{W}_5]$
is not effective. Instead, we find the following minimal solutions,
depending on whether one works in the strongly or the weakly
coupled regime of the heterotic string.

\subsubsection {Strong Coupling: Bulk Five-branes}

The minimal
vector bundle $\Vt'$ that is consistent with anomaly constraint~\eqref{eff}
is found to have structure group $SU(2)$. For this bundle,
\begin{equation}
[\mathcal{W}_5] \neq  0 
\label{ii1}
\end{equation}
and, hence, this hidden sector is compatible only with the strongly 
coupled heterotic string. $\Vt'$ spontaneously breaks the hidden sector $E_8'$ 
gauge symmetry to
\begin{equation}
E_{8}' \longrightarrow E_{7}.
\label{ii2}
\end{equation}
With respect to $SU(2) \times E_7$, the adjoint
representation of $E_8'$ decomposes as
\begin{equation}
\Rep{248}'=
\big( \Rep{1}, \Rep{133} \big) \oplus
\big( \Rep{3}, \Rep{1}   \big) \oplus
\big( \Rep{2}, \Rep{56}  \big)
\,.
\end{equation}
The $\big( \Rep{1}, \Rep{133} \big)$ contain the gauginos of $E_{7}$,
the $\big( \Rep{3}, \Rep{1}   \big)$ correspond to vector bundle moduli and 
$\big( \Rep{2}, \Rep{56}  \big)$ represent charged exotic matter fields.  
In addition to demanding that $\Vt'$ satisfy the
stability conditions~\eqref{stability}, we require that there be 
no exotic matter in the hidden sector. This is most easily accomplished by
imposing the constraint that
\begin{equation}
  h^1( \Xt, \Vt')=0
  \,.
\label{jj1}
\end{equation}
The requisite $SU(2)$ bundle $\Vt'$ is any element of the space of
extensions
\begin{equation}\label{H}
\sseq{\cO_{\Xt}(2\tau_1+\tau_2-\phi)}{\Vt'}{\cO_{\Xt}(-2\tau_1-\tau_2+\phi)}.
\end{equation}
One can easily show that the entire cohomology ring vanishes. 
That is 
\begin{equation}\label{HH} h^i(\Xt, \Vt') = 0, \qquad
i=0,1,2,3. 
\end{equation}
Note that this result is consistent with the necessary conditions~\eqref{stability} that 
$\Vt'$ be stable. Furthermore, it follows that~\eqref{jj1} is satisfied 
and, hence, there is no exotic matter in the hidden sector.

The five-brane wrapped on a holomorphic curve associated with
$[\mathcal{W}_5]$ contributes non-Abelian gauge fields, but no matter
fields, to the hidden sector. Following the results
in~\cite{Lukas:1999kt, Lukas:1998uy}, we find that the five-brane
gauge group is $U(6)$.  Moving in the moduli space of the holomorphic
curve, this group can be maximally broken to $U(1)^6$.

We conclude that, within the context of the strongly coupled
heterotic string, our observable sector is consistent with a hidden
sector 1) with gauge group $E_7 \times U(6)$ and 2) {\it no exotic
matter}. In addition, 3) there is a {\it small number} of vector 
bundle moduli arising from the invariant subspace of $H^{1}(\Xt, \Vt' \otimes
{\Vt}^{'*})$ under the action of $\z3z3$, as well as some five-brane moduli. 
These will be computed elsewhere.

\subsubsection{Weak Coupling: No Five-branes}

We now exhibit a hidden sector, consistent with our observable
sector, that has no five-branes; that is, for which
\begin{equation}\label{W=01}
  [\mathcal{W}_5]=0.
\end{equation}
This hidden sector is compatible with both the weakly and strongly
coupled heterotic string. We are unable to satisfy~\eqref{W=01} for
any bundle with an $SU(2)$ structure group. From the results
in~\cite{Buchbinder:2002ji, He:2003tj}, we expect that the appropriate
structure group may be the product of two non-Abelian groups, the
simplest choice being $SU(2) \times SU(2)$.  This bundle, which is the
sum of two $SU(2)$ factors, $\Vt'=\Vt_1' \oplus \Vt_2'$, spontaneously
breaks
\begin{equation}
  E_8' \longrightarrow Spin(12).
\label{iia}
\end{equation}
With respect to $SU(2) \times SU(2) \times
Spin(12)$, the adjoint representation of $E_8'$ decomposes as
\begin{equation}
\Rep{248}'= \big(\Rep{1},\Rep{1},\Rep{66} \big) \oplus
\big(\Rep{3}, \Rep{1}, \Rep{1} \big) \oplus
\big(\Rep{1}, \Rep{3}, \Rep{1} \big) \oplus \big(\Rep{2}, \Rep{1}, 
\Rep{32} \big)
\oplus \big(\Rep{1}, \Rep{2}, \Rep{32} \big) \oplus \big(\Rep{2},
\Rep{2}, \Rep{12} \big) \,. \label{32}
\end{equation}
Representations $\big(\Rep{1},\Rep{1},\Rep{66} \big) $ and 
$\big(\Rep{3}, \Rep{1}, \Rep{1} \big) \oplus
\big(\Rep{1}, \Rep{3}, \Rep{1} \big)$ contain the $Spin(12)$ gauginos 
and vector bundle moduli.
Exotic matter in the hidden sector can arise from 
$\big(\Rep{2}, \Rep{1}, \Rep{32} \big)$, $\big(\Rep{1}, \Rep{2},
\Rep{32} \big)$, and $\big(\Rep{2}, \Rep{2}, \Rep{12} \big)$,
corresponding to the cohomology spaces $H^1( \Xt,
\Vt_1' )$, $H^1( \Xt, \Vt_2' )$, and $H^1( \Xt, \Vt_1' \otimes
\Vt_2' )$ respectively. Unlike the case in the strong coupling
regime, subject to \eqref{W=01} and the stability
conditions \eref{stability} applied to $\Vt'_{1,2}$, we are unable
to find a hidden sector bundle for which all
exotic matter is absent. 

However, relaxing the constraints so that a small amount of hidden
exotic matter may exist, one finds the following minimal solution.
It turns out that $\Vt'_1$ is the bundle $\Vt'$ introduced in
\eref{H} and $\Vt'_2$ is the pullback of an extension on $B_1$. Specifically,
$\Vt'_2 = \pi_1^*\cS_B$, where
\begin{equation}
    \sseq{\cO_{B_1}(-2f_{1})}{\cS_B}{\cO_{B_1}(2f_{1}) \otimes I_6}.
\end{equation}
Here, $I_6$ is the ideal sheaf of 6 points on $B_1$ which are a
single orbit of $g_2 \in \z3z3$ with multiplicity 2.

Recall from \eref{HH} that $h^i(\Xt, \Vt_1')$ for $i=0,1,2,3$ vanish. 
Therefore, $\Vt_{1}'$ satisfies the stability conditions~\eqref{stability}
and there is no matter in the $\big(\Rep{2}, \Rep{1}, \Rep{32} \big)$
representation. For $\Vt_2'$, we find that
%
\begin{equation}
h^0(\Xt, \Vt_2')=h^3(\Xt, \Vt_2')=0, \quad h^1(\Xt, \Vt_2')=h^2(\Xt, \Vt_2')=4.
\label{lll1}
\end{equation}
Furthermore, for $\Vt_1' \otimes \Vt_2'$ one can show 
\begin{equation}
h^0\big(\Xt, \Vt_1' \otimes \Vt_2'\big)=
h^3\big(\Xt, \Vt_1' \otimes \Vt_2'\big)=0
\label{xx1}
\end{equation}
and
\begin{equation}
h^1\big(\Xt, \Vt_1' \otimes \Vt_2'\big)=h^2\big(\Xt, \Vt_1' \otimes 
\Vt_2'\big)=18.
\label{m1}
\end{equation}
%
If follows from~\eqref{lll1} that $\Vt_{2}'$ also satisfies the stability
constraints~\eqref{stability}. However,
$h^1(\Xt, \Vt_2')$ and $h^1\big(\Xt, \Vt_1' \otimes
\Vt_2'\big)$ do not vanish and may give rise to hidden sector exotic matter
in the representations $\big(\Rep{1}, \Rep{2},
\Rep{32} \big)$ and $\big(\Rep{2}, \Rep{2}, \Rep{12} \big)$ respectively.

To analyze this, it is necessary to explicitly compute the 
action of $\z3z3$ on these cohomology spaces. This can be accomplished using
methods similar to those discussed previously. 
Here, we simply state the results. The action of $\z3z3$ on
$H^1(\Xt, \Vt_1')$ and $H^1\big(\Xt, \Vt_1' \otimes \Vt_2'\big)$ is 
found to be
\begin{equation}
H^1(\Xt, \Vt_1')= 2\chi_1 \oplus 2\chi_1^2
\label{lll2}
\end{equation}
and 
\begin{equation}
H^1\big(\Xt, \Vt_1' \otimes \Vt_2'\big)=  {\rm Reg}(\z3z3)^{\oplus 2}
\label{lll3}
\end{equation}
respectively. It follows from~\eqref{lll2} that 
$H^1\big(\Xt, \Vt_2'\big)$ has no 
invariant subspace. Since there is no Wilson line in the 
hidden sector, 
all $\big(\Rep{1}, \Rep{2}, \Rep{32} \big)$ exotic matter
fields are projected out of the low energy spectrum. Unfortunately, this is 
not the case for $H^1\big(\Xt, \Vt_1' \otimes \Vt_2'\big)$. Action~\eqref{lll3}
implies that there remain two exotic $\Rep{12}$ multiplets of
$Spin(12)$ after projection. 

We conclude that, for vacua with no five-branes, our observable
sector is consistent with a hidden sector 1) with gauge group
$Spin(12)$ and 2) {\it two} $\Rep{12}$ multiplets. We emphasize 
that these are not charged under the observable sector gauge group.
There are also vector bundle moduli
arising from the $\Z_3 \times \Z_3$ invariant subspace of
$H^1(\Xt, \Vt' \otimes \Vt^{'*})$, which will be computed elsewhere. 
These vacua can occur in the context of both the weakly and strongly 
coupled heterotic string.


\section*{Acknowledgments}
We are grateful to P.~Candelas, R.~Donagi, P.~Langacker, B.~Nelson,
R.~Reinbacher and D.~Waldram for enlightening discussions. We also
thank L.~Iba{\~n}ez, H.P.~Nilles, G.~Shiu and A.~Uranga for helpful
conversations. This research was supported in part by the Department
of Physics and the Math/Physics Research Group at the University of
Pennsylvania under cooperative research agreement DE-FG02-95ER40893
with the U.~S.~Department of Energy and an NSF Focused Research Grant
DMS0139799 for ``The Geometry of Superstrings.'' T.~P.~is partially
supported by an NSF grant DMS 0104354 and DMS 0403884.


\bibliographystyle{JHEP} \renewcommand{\refname}{Bibliography}
\addcontentsline{toc}{section}{Bibliography} 
{
  \renewcommand{\baselinestretch}{1}
  \bibliography{main}

\providecommand{\href}[2]{#2}\begingroup\raggedright\begin{thebibliography}{10}

\bibitem{VBphysicsletter}
V.~Braun, Y.-H. He, B.~A. Ovrut, and T.~Pantev, {\it A heterotic standard
  model},  \href{http://xxx.lanl.gov/abs/hep-th/0501070}{{\tt hep-th/0501070}}.

\bibitem{Witten:1995gx}
E.~Witten, {\it Small instantons in string theory},  {\em Nucl. Phys.} {\bf
  B460} (1996) 541--559, [\href{http://xxx.lanl.gov/abs/hep-th/9511030}{{\tt
  hep-th/9511030}}].

\bibitem{Curio:1998vu}
G.~Curio, {\it Chiral matter and transitions in heterotic string models},  {\em
  Phys. Lett.} {\bf B435} (1998) 39--48,
  [\href{http://xxx.lanl.gov/abs/hep-th/9803224}{{\tt hep-th/9803224}}].

\bibitem{Ovrut:2000qi}
B.~A. Ovrut, T.~Pantev, and J.~Park, {\it Small instanton transitions in
  heterotic {M}-theory},  {\em JHEP} {\bf 05} (2000) 045,
  [\href{http://xxx.lanl.gov/abs/hep-th/0001133}{{\tt hep-th/0001133}}].

\bibitem{dP9Z3Z3}
V.~Braun, B.~A. Ovrut, T.~Pantev, and R.~Reinbacher, {\it Elliptic {Calabi-Yau}
  threefolds with {$\Z_3 \times \Z_3$} {W}ilson lines},  {\em JHEP} {\bf 12}
  (2004) 062, [\href{http://xxx.lanl.gov/abs/hep-th/0410055}{{\tt
  hep-th/0410055}}].

\bibitem{FMW}
R.~Friedman, J.~Morgan, and E.~Witten, {\it Vector bundles and {F} theory},
  {\em Commun. Math. Phys.} {\bf 187} (1997) 679--743,
  [\href{http://xxx.lanl.gov/abs/hep-th/9701162}{{\tt hep-th/9701162}}].

\bibitem{Friedman:1997ih}
R.~Friedman, J.~W. Morgan, and E.~Witten, {\it Vector bundles over elliptic
  fibrations},  \href{http://xxx.lanl.gov/abs/alg-geom/9709029}{{\tt
  alg-geom/9709029}}.

\bibitem{Andreas:1999ty}
B.~Andreas, G.~Curio, and A.~Klemm, {\it Towards the standard model spectrum
  from elliptic {Calabi- Yau}},  {\em Int. J. Mod. Phys.} {\bf A19} (2004)
  1987, [\href{http://xxx.lanl.gov/abs/hep-th/9903052}{{\tt hep-th/9903052}}].

\bibitem{DonagiPrincipal}
R.~Y. Donagi, {\it Principal bundles on elliptic fibrations},  {\em Asian J.
  Math.} {\bf 1} (1997), no.~2 214--223,
  [\href{http://xxx.lanl.gov/abs/alg-geom/9702002}{{\tt alg-geom/9702002}}].

\bibitem{Donagi:1998xe}
R.~Donagi, A.~Lukas, B.~A. Ovrut, and D.~Waldram, {\it Non-perturbative vacua
  and particle physics in {M}-theory},  {\em JHEP} {\bf 05} (1999) 018,
  [\href{http://xxx.lanl.gov/abs/hep-th/9811168}{{\tt hep-th/9811168}}].

\bibitem{Donagi:1999gc}
R.~Donagi, A.~Lukas, B.~A. Ovrut, and D.~Waldram, {\it Holomorphic vector
  bundles and non-perturbative vacua in {M}- theory},  {\em JHEP} {\bf 06}
  (1999) 034, [\href{http://xxx.lanl.gov/abs/hep-th/9901009}{{\tt
  hep-th/9901009}}].

\bibitem{Curio:2004pf}
G.~Curio, {\it Standard model bundles of the heterotic string},
  \href{http://xxx.lanl.gov/abs/hep-th/0412182}{{\tt hep-th/0412182}}.

\bibitem{Andreas:2003zb}
B.~Andreas and D.~Hernandez-Ruiperez, {\it Comments on {$N=1$} heterotic string
  vacua},  {\em Adv. Theor. Math. Phys.} {\bf 7} (2004) 751--786,
  [\href{http://xxx.lanl.gov/abs/hep-th/0305123}{{\tt hep-th/0305123}}].

\bibitem{Diaconescu:1998kg}
D.-E. Diaconescu and G.~Ionesei, {\it Spectral covers, charged matter and
  bundle cohomology},  {\em JHEP} {\bf 12} (1998) 001,
  [\href{http://xxx.lanl.gov/abs/hep-th/9811129}{{\tt hep-th/9811129}}].

\bibitem{Donagi:2000fw}
R.~Donagi, B.~A. Ovrut, T.~Pantev, and D.~Waldram, {\it Spectral involutions on
  rational elliptic surfaces},  {\em Adv. Theor. Math. Phys.} {\bf 5} (2002)
  499--561, [\href{http://xxx.lanl.gov/abs/math.ag/0008011}{{\tt
  math.ag/0008011}}].

\bibitem{Donagi:2000zs}
R.~Donagi, B.~A. Ovrut, T.~Pantev, and D.~Waldram, {\it Standard-model
  bundles},  {\em Adv. Theor. Math. Phys.} {\bf 5} (2002) 563--615,
  [\href{http://xxx.lanl.gov/abs/math.ag/0008010}{{\tt math.ag/0008010}}].

\bibitem{Donagi:2000zf}
R.~Donagi, B.~A. Ovrut, T.~Pantev, and D.~Waldram, {\it Standard-model bundles
  on non-simply connected {Calabi-Yau} threefolds},  {\em JHEP} {\bf 08} (2001)
  053, [\href{http://xxx.lanl.gov/abs/hep-th/0008008}{{\tt hep-th/0008008}}].

\bibitem{Donagi:1999ez}
R.~Donagi, B.~A. Ovrut, T.~Pantev, and D.~Waldram, {\it Standard models from
  heterotic {M}-theory},  {\em Adv. Theor. Math. Phys.} {\bf 5} (2002) 93--137,
  [\href{http://xxx.lanl.gov/abs/hep-th/9912208}{{\tt hep-th/9912208}}].

\bibitem{Donagi:2003tb}
R.~Donagi, B.~A. Ovrut, T.~Pantev, and R.~Reinbacher, {\it {$SU(4)$} instantons
  on {Calabi-Yau} threefolds with {$\Z_2\times\Z_2$} fundamental group},  {\em
  JHEP} {\bf 01} (2004) 022,
  [\href{http://xxx.lanl.gov/abs/hep-th/0307273}{{\tt hep-th/0307273}}].

\bibitem{Ovrut:2003zj}
B.~A. Ovrut, T.~Pantev, and R.~Reinbacher, {\it Invariant homology on standard
  model manifolds},  {\em JHEP} {\bf 01} (2004) 059,
  [\href{http://xxx.lanl.gov/abs/hep-th/0303020}{{\tt hep-th/0303020}}].

\bibitem{Ovrut:2002jk}
B.~A. Ovrut, T.~Pantev, and R.~Reinbacher, {\it Torus-fibered {Calabi-Yau}
  threefolds with non-trivial fundamental group},  {\em JHEP} {\bf 05} (2003)
  040, [\href{http://xxx.lanl.gov/abs/hep-th/0212221}{{\tt hep-th/0212221}}].

\bibitem{Donagi:2004su}
R.~Donagi, Y.-H. He, B.~A. Ovrut, and R.~Reinbacher, {\it Higgs doublets, split
  multiplets and heterotic {$SU(3)_C \times SU(2)_L \times U(1)_Y$} spectra},
  \href{http://xxx.lanl.gov/abs/hep-th/0409291}{{\tt hep-th/0409291}}.

\bibitem{Donagi:2004ub}
R.~Donagi, Y.-H. He, B.~A. Ovrut, and R.~Reinbacher, {\it The spectra of
  heterotic standard model vacua},
  \href{http://xxx.lanl.gov/abs/hep-th/0411156}{{\tt hep-th/0411156}}.

\bibitem{Donagi:2004qk}
R.~Donagi, Y.-H. He, B.~A. Ovrut, and R.~Reinbacher, {\it Moduli dependent
  spectra of heterotic compactifications},
  \href{http://xxx.lanl.gov/abs/hep-th/0403291}{{\tt hep-th/0403291}}.

\bibitem{Donagi:2004ia}
R.~Donagi, Y.-H. He, B.~A. Ovrut, and R.~Reinbacher, {\it The particle spectrum
  of heterotic compactifications},
  \href{http://xxx.lanl.gov/abs/hep-th/0405014}{{\tt hep-th/0405014}}.

\bibitem{Green:1987mn}
M.~B. Green, J.~H. Schwarz, and E.~Witten, {\it Superstring theory. vol. 2:
  Loop amplitudes, anomalies and phenomenology}, . Cambridge, Uk: Univ. Pr. (
  1987) 596 P. ( Cambridge Monographs On Mathematical Physics).

\bibitem{Witten:2001bf}
E.~Witten, {\it Deconstruction, {$G_2$} holonomy, and doublet-triplet
  splitting},  \href{http://xxx.lanl.gov/abs/hep-ph/0201018}{{\tt
  hep-ph/0201018}}.

\bibitem{VBpaper}
V.~Braun, Y.-H. He, B.~A. Ovrut, and T.~Pantev, ``Holomorphic bundles and
  standard-model vacua in heterotic string theory.'' To appear.

\bibitem{multipletsplitting}
V.~Braun, Y.-H. He, B.~A. Ovrut, and T.~Pantev, ``Doublet-triplet splitting
  from heterotic string theory.'' To appear.

\bibitem{Fukuda:1998mi}
{\bf Super-Kamiokande} Collaboration, Y.~Fukuda {\em et.~al.}, {\it Evidence
  for oscillation of atmospheric neutrinos},  {\em Phys. Rev. Lett.} {\bf 81}
  (1998) 1562--1567, [\href{http://xxx.lanl.gov/abs/hep-ex/9807003}{{\tt
  hep-ex/9807003}}].

\bibitem{Langacker:2004xy}
P.~Langacker, {\it Neutrino physics (theory)},
  \href{http://xxx.lanl.gov/abs/hep-ph/0411116}{{\tt hep-ph/0411116}}.

\bibitem{Giedt:2005vx}
J.~Giedt, G.~L. Kane, P.~Langacker, and B.~D. Nelson, {\it Massive neutrinos
  and (heterotic) string theory},
  \href{http://xxx.lanl.gov/abs/hep-th/0502032}{{\tt hep-th/0502032}}.

\bibitem{Candelas:1985en}
P.~Candelas, G.~T. Horowitz, A.~Strominger, and E.~Witten, {\it Vacuum
  configurations for superstrings},  {\em Nucl. Phys.} {\bf B258} (1985)
  46--74.

\bibitem{Sen:1985eb}
A.~Sen, {\it The heterotic string in arbitrary background field},  {\em Phys.
  Rev.} {\bf D32} (1985) 2102.

\bibitem{Witten:1985xc}
E.~Witten, {\it Symmetry breaking patterns in superstring models},  {\em Nucl.
  Phys.} {\bf B258} (1985) 75.

\bibitem{Breit:1985ud}
J.~D. Breit, B.~A. Ovrut, and G.~C. Segre, {\it {$E(6)$} symmetry breaking in
  the superstring theory},  {\em Phys. Lett.} {\bf B158} (1985) 33.

\bibitem{Greene:1986ar}
B.~R. Greene, K.~H. Kirklin, P.~J. Miron, and G.~G. Ross, {\it A superstring
  inspired standard model},  {\em Phys. Lett.} {\bf B180} (1986) 69.

\bibitem{Greene:1986jb}
B.~R. Greene, K.~H. Kirklin, P.~J. Miron, and G.~G. Ross, {\it A three
  generation superstring model. 2. symmetry breaking and the low-energy
  theory},  {\em Nucl. Phys.} {\bf B292} (1987) 606.

\bibitem{Aspinwall:1987cn}
P.~S. Aspinwall, B.~R. Greene, K.~H. Kirklin, and P.~J. Miron, {\it Searching
  for three generation {C}alabi-{Y}au manifolds},  {\em Nucl. Phys.} {\bf B294}
  (1987) 193.

\bibitem{Lukas:1999kt}
A.~Lukas, B.~A. Ovrut, and D.~Waldram, {\it Five-branes and supersymmetry
  breaking in {M}-theory},  {\em JHEP} {\bf 04} (1999) 009,
  [\href{http://xxx.lanl.gov/abs/hep-th/9901017}{{\tt hep-th/9901017}}].

\bibitem{Lukas:1998uy}
A.~Lukas, B.~A. Ovrut, and D.~Waldram, {\it Heterotic {M}-theory vacua with
  five-branes},  {\em Fortsch. Phys.} {\bf 48} (2000) 167--170,
  [\href{http://xxx.lanl.gov/abs/hep-th/9903144}{{\tt hep-th/9903144}}].

\bibitem{Buchbinder:2002ji}
E.~Buchbinder, R.~Donagi, and B.~A. Ovrut, {\it Vector bundle moduli and small
  instanton transitions},  {\em JHEP} {\bf 06} (2002) 054,
  [\href{http://xxx.lanl.gov/abs/hep-th/0202084}{{\tt hep-th/0202084}}].

\bibitem{He:2003tj}
Y.-H. He, B.~A. Ovrut, and R.~Reinbacher, {\it The moduli of reducible vector
  bundles},  {\em JHEP} {\bf 03} (2004) 043,
  [\href{http://xxx.lanl.gov/abs/hep-th/0306121}{{\tt hep-th/0306121}}].

\end{thebibliography}\endgroup
}

\end{document}